# OPEN INPUT: A NEW WAY FOR WEBSITES TO GROW

Pavel Vasev

*LineAct, Ekaterinburg, Russian Federation*
contact@lact.ru

**ABSTRACT**

Regardless of current web 2.0 and 3.0 trends, there are still a lot of websites made in web 1.0 style. These websites have fixed pages which are editable only by owner and not by community. It is normal for a lot of cases, but looks like not modern and engaging approach. Are there any ways to make these sites closer to life? This paper is devoted to *open input* technique, a way for websites of web 1.0 era to grow and evolve community. The idea of open input, in general, means that anybody from the web can add information to any section of the website even without registration on that website. People can add news, billboard announcements, testimonials, questions, pictures, videos etc – whatever site owner permitted. We have tested this idea in practice and have positive results approving that open input is a vital approach for collaboration on the web.

## 1. INTRODUCTION. THE PROBLEM.

Nowadays the collaboration is a mainstream trend on the web [O'Relly]. A lot of organizations use tools such as Facebook, Twitter and MySpace to provide and maintain interaction arena for users and between users. These tools are used in addition to traditional websites. Why website owners prefer to use Facebook etc for collaboration with people? Because of its large existing user base and because involving is simple: user has just to click "join". Another reason why website owners prefer external tools is that existing content management systems (CMS) of websites are poorly adopted for collaboration. Thus, most CMS have forums, collective blogs, web forms, etc. However, if website owner wants to add new collaboration tool, he has to ask CMS developer to implement these tool. So every time a new idea appears, it should be implemented by developer. Create a billboard announcements controlled by users feature? No problem, 2 weeks of work. Add testimonials gathering using web forms and transparently publish them? No problem, 2 more weeks of work. And so on and so on. It is natural behavior for CMS developer: he just cannot foreknow all features of user interaction nor implement them because the number of such features is countless. So website owner should waste time and money for implementation of each new feature again and again. This is the problem which prevents website owners with limited budget from creating communities based on their own websites.

Are there any ways to automate the creation of collaboration tools for websites? If such a way exists, it could help to improve interaction on websites because rolling out new collaboration channels will be faster and cheaper. It is well known that in general any problem has universal solution in limited bounds. We hope that our case is not an exception and we will introduce the one method of automating the process of creation of collaboration tools for websites below in this article.

## 2. THE SOLUTION

We may start by examining the main compounding parts of web site. In general, these parts are sections and content elements. Website sections are structural items which form website tree. For example, in typical business's website sections are Products, Services, Contacts, Testimonials, etc. Content elements are items which contain actual information. Content elements are tied into sections. In our example, content element may be a product item, one contact address, one testimonial, etc. Website owner creates one or more sections of appropriate type, adds description for them and then adds one or more content elements of according types. Of course not all websites are constructed in this way, but a lot of them are. Terminology may vary, for example content elements may be called pages, or items, or whatever developer named them. In our case at research project called LineAct CMS [Vasev] the things behave exactly as stated here.

In common, at almost all websites on the Internet, only the site owner (or another authorized person) may add elements into sections. Now let's imagine: what if we will change this and allow everybody to add information in specified sections? According to this approach, site owner configures for which sections on website he want to open public access. He also specifies element types which may be added to the section. For example, he may restrict system so only testimonials may be added into the Testimonials section of his website, and only textual elements may be added into the Billboard section. Finally, website owner specifies who is allowed to add information: unauthenticated visitors, open-id authenticated or only website registered users, etc.

When a website visitor wants to add some information to that website, he clicks "Add information" link and fills out exactly the same input web form of CMS as website owner during information input process. Upon completion, entered data goes to confirmation queue and website owner receives notification. Then he passes through that queue, examines user-generated information and allows it to be published on website or declines it. Of course, trusted users may omit confirmation process, but for non-authenticated visitors access confirmation queue is mandatory.

We call the technique described above as *open input*. It covers a lot of web collaboration cases. Website owners may configure their websites to do a lot of new things using following algorithm: gather information of specified type for website by one click and then publish it on website, also by one click. Of course user-entered information nests all necessary attributes: it may be rated, commented, sorted, etc. Using open input technique, a website owner may implement, for example, collective blog or newsfeed, testimonials gathering, public billboard, collective recipes, questions and answers section, user photo ratings, and so on.

The main benefit of the open input is that website owner is not restricted to element information type and also may turn on information gathering on for any desired section of his website. The feature may be turned on whenever needed with no developer skills required.

The second benefit of open input technique is that it may work even for non-authenticated users because of confirmation queue. It is a fully secure mechanism: no unwanted information will be published. Working with non-authenticated users is a very important aspect of website: it involves people to collaborate without asking anything from them. Website visitor is engaged into collaboration at "no cost" and very fast by just one click. If he founds collaboration on this website useful, he may become a registered user of that website. This will happen with higher level of probability opposite to mandatory registration scenario.

The third benefit of the open input is that it may be implemented in any CMS at very low cost level. Actually, CMS developers do not need to implement special input forms for public open input – original administrator CMS input forms may be used for open input.

## 3. RELATED WORKS

There are a lot of solutions for people collaboration and involvement on the Web. These solutions rely on various communication models. We must note and provide short comment for at least the following:
- Wiki systems – allows people to publish pages into global structured information space.
- Social blogging systems – less flexible but more structured way to publish information and receive feedback.
- Forums – most flexible and most unstructured way for publishing and discussion processes.
- Billboards, social news systems, aggregation systems.

The main differences of open input from already existing approaches stated above are:
1. It has as simpler interaction schema as it possible. User just has to click "Add information" and then enter title, textual and related content. No more steps required.
2. It does not require user registration. It is very important usability gain [Cooper].
3. It is very flexible: using open input, billboards, forums, social news and other models may be implemented. Of course, such flexibility lacks structure, but it is price of the flexibility.

4. It is easy to implement and embed into existing website infrastructure.

## 4. THE EXPERIMENT

The stated open input technique was implemented and introduced in LineAct CMS at August 2009. LineAct is a simple CMS system and it has very few collaboration tools build-in, and even still has no user registration capabilities. To achieve open input behavior, the following changes were made:
1. Added "status" field into DB schema for every information on the website. Possible values are: "pending", "declined", "accepted".
2. Added an ability to show button titled "Add your information" onto every section of website.

The system acts by the following way. When website owner add information to website, it always become "accepted". When some website visitor clicks "add your information" button, he fulfills the same CMS webform as website owner, but added information receives "pending" status. Nor pending nor declined information are never shown on website's public area – only website owner can see it. So when visitor adds information, owner receives email notification attending him to perform moderation on incoming pending queue. Owner looks over new information listing and clicks AJAX-styled resolution links "accept" or "decline". Because of its simple interaction method (the decision is made by just one click without page reload), the moderation takes 5-10 minutes in a day for even small to medium-sized websites.

We can see an example of website with open input technique at the following figure 1. This is koni66.ru portal dedicated to horse clubs of Sverdlovsk area. The figure shows us News section of that website. Note the "Click here to add your information" button in the top part of the figure. This is an entry point into open input approach. Thanks to this feature, all horse clubs' representatives may post their news into that website.

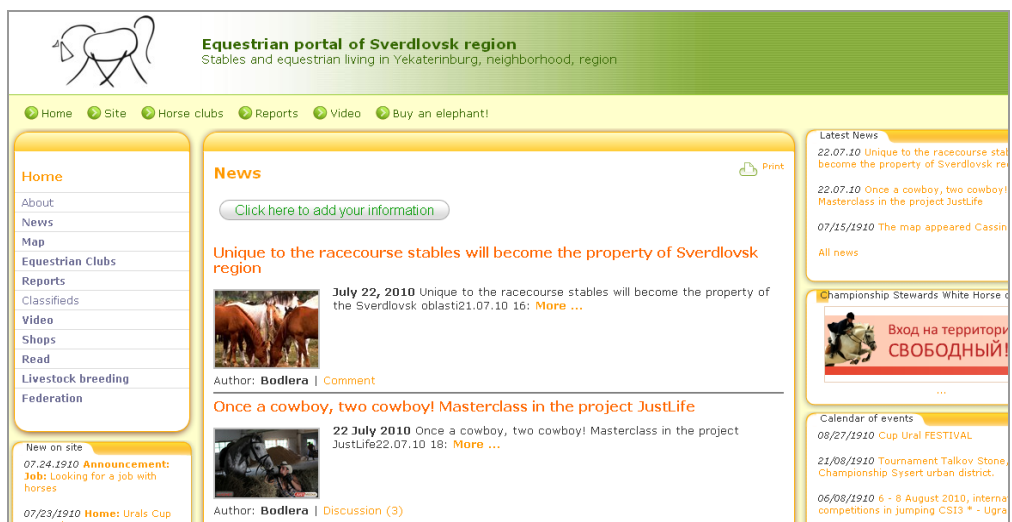

Figure 1. An example of website with embedded open input technique.

The importance of this example is higher due to the fact that a lot of horse club's owners even have no email addresses at all! And of course, they do not have their own websites etc. So their only ability to be presented on the Web is to post information onto shown website. This works well because our open input implementation doesn't require nor registration nor emails from publishing users. Actually, the behavior is the following:
2. If the user provides email during information publish process, the system generates a special "editor-link" with randomly generated GUID bound to his email address and sends this link by email. Having such a link, the user is able to edit or delete published information later.
3. By the other hand, if no email were provided, no editor-link is generated, but the publishing process continues as usual. However, such link may be generated later by website owner upon request.

So we described the idea and sample implementation of open input technique. Let's consider some statistics gathered during 9 month period after this technique was introduced to public.

- Open input was used on 526 websites in total. Because LineAct is a paid hosted service, only 128 of these websites are "live" (not closed) now, and the total number of live websites is 387.
- Totally 7226 content elements of various semantic types were entered by users.
- 4061 of this amount were accepted by website owners and thus published.
- Most of these elements – 3149 – belong to only top 41 websites.

The semantic types of published information are various. These types are divided into two classes: built-in and custom. Users of LineAct had published following information of built-in semantic types:
- 274 client testimonials.
- 705 billboard announcements.
- 560 questions and answers.
- 43 news and events.
- 65 firm client information.

Beside build-in semantic types, there are a number of custom information types in LineAct CMS, for example: text, video, link, image gallery. Each of this custom type may be used in many semantic means. Because of large numbers of information entered (for example, users published 559 text elements), we still have no ability to analyze it all. However first analyzes shows us that users had used custom types by the following means: company contacts (for example, they entered contacts of horse clubs at horse's portal figured above), forum messages, contest paintings, etc. Actually, we need more investigation to find all the ways that website visitors use open input for. The main idea of such analysis is to find new interaction models and detect absent built-in semantic types for CMS. At the moment, we have no final results on it.

## 5. CONCLUSION

We consider the open input technique as one of the universal ways for websites to perform collaboration with its users. Our considerations are based on the experiment that was started in August 2009. Currently, after 9 month from introduction, the open input is used in 128 websites out of total 387 active websites running LineAct CMS. These numbers shows us that proposed approach became popular. We conclude that open input technique is vital and may be used in practice.